\documentclass[9pt,twocolumn,twoside]{myclass}

\usepackage{comment}
\usepackage{xr}
\usepackage{cleveref}

\title{{\Huge Diffractive optical system design by cascaded propagation}}

\author[1,2]{Boris Ferdman}
\author[2]{Alon Saguy}
\author[2]{Onit Alalouf}
\author[1,2,*]{Yoav Shechtman}

\affil[1]{Russel Berrie Nanotechnology Institute, Technion - Israel Institute of Technology, 3200003 Haifa, Israel}
\affil[2]{Department of Biomedical Engineering \& Lorry I. Lokey Center for Life Sciences and Engineering,\linebreak Technion - Israel Institute of Technology, 3200003 Haifa, Israel}

\affil[*]{Corresponding author: \href{mailto:yoavsh@bm.technion.ac.il}{yoavsh@bm.technion.ac.il}}

\begin{document}
\maketitle

\begin{abstract}

Modern design of complex optical systems relies heavily on computational tools. These typically utilize geometrical optics as well as Fourier optics, which enables the use of diffractive elements to manipulate light with features on the scale of a wavelength. Fourier optics is typically used for designing thin elements, placed in the system's aperture, generating a shift-invariant Point Spread Function (PSF). A major bottleneck in applying Fourier Optics in many cases of interest, \emph{e.g.} when dealing with multiple, or out-of-aperture elements, comes from numerical complexity. In this work, we propose and implement an efficient and differentiable propagation model based on the Collins integral, which enables the optimization of diffraction optical systems with unprecedented design freedom using backpropagation. We demonstrate the applicability of our method, numerically and experimentally, by engineering shift-variant PSFs \emph{via} thin plate elements placed in arbitrary planes inside complex imaging systems, performing cascaded optimization of multiple planes, and designing optimal machine-vision systems by deep learning.

\end{abstract}

\maketitle

\section{Introduction} \label{sec:intro}

Designing optical imaging systems can be a challenging task, addressable by various different approaches, depending on the specifics of the problem. The  commonly used geometrical optics (ray tracing) method is a crude approximation of wave optics, yet it provides powerful tools to design thick elements and complex stacks of lenses. However, geometrical optics is not suitable for applications such as phase retrieval \cite{fienup1982phase,millane1990phase,shechtman2015phase,yeh2015experimental}, adaptive optics \cite{ji2017adaptive,hampson2021adaptive}, PSF engineering \cite{huang2008three,jesacher2016colored,shechtman2014optimal,elmalem2020motion}, or more generally, any application that requires either correctly modelling the diffraction-limited PSF or employing optical elements that manipulate the phase, amplitude, or polarization of light. Such applications require Fourier optics \cite{goodman2005introduction}, which can account for diffraction effects, but is usually limited to thin-element and paraxial approximations. Today, a prevalent approach \cite{born2013principles} is to use a hybrid model, where first ray-tracing is employed to efficiently compute the Optical-Path Difference (OPD) from the object space to the exit pupil. Next, a diffraction model, based on Fourier optics, is used to compute the 2D/3D Point-Spread Function (PSF) on the detector plane. Existing optical engineering tools  (\emph{e.g.} Zemax \cite{zemax}, CodeV \cite{CodeV}, \emph{etc.}) can then be used to optimize the optical components by numerically minimizing a hand-crafted merit function. This approach takes into account the thickness of optical elements, spatially variant aberrations, dispersion effects and more.

Fourier optics is exceptionally useful in the field of computational imaging, where optical systems are co-designed alongside with analysis algorithms, yielding powerful imaging systems with non-conventional capabilities; applications range from High Dynamic Range imaging \cite{alghamdi2021transfer,sun2020learning}, Color sensing \cite{baek2021single,arguello2021shift}, to end-to-end wavefront shaping \cite{tan2021codedstereo,nehme2020deepstorm3d,wu2020freecam3d} and more. However, because of their reliance on Fourier optics, these applications have traditionally been limited to the optimization of a single optical element, typically placed in the system's aperture. Recently, alternative end-to-end computational photography  methods based on ray tracing have been proposed \cite{li2021end,tseng2021differentiable,sun2021end}, showing promise in computer vision applications. These methods optimize lens stack and can do polynomial surfaces, but cannot utilize the effect of diffraction. Other techniques which use ray tracing-based Monte Carlo methods \cite{arasa2004comparison} exist and can be suitable for specific challenging cases (such as very short propagation distances) but are not suitable for direct optimization of optical systems.   

One applicable approach for analyzing and designing optical systems with simple components is using the Collins integral \cite{collins1970lens,li2008algorithm}, a parabolic phase case of the Fractional Fourier transform \cite{mendlovic1993fractional,su2019analysis}, which facilitates an angular-spectrum-based method to compute the diffraction integral using the ray transfer matrix. Recently, a cascaded diffraction model based on the Collins integral was implemented to model the diffraction of a series of diaphragms in complex optical systems \cite{gross2020cascaded,gross2020cascaded2} to study and accurately simulate spatially variant edge-diffraction effects. Spatially-variant PSFs are usually digitally corrected as they can degrade the performance of algorithms which rely on a spatially-invariant PSF computation \cite{shajkofci2020spatially,debarnot2021deepblur}, while designing spatially-variant PSFs (in Fourier optics) is considered in some specific cases such as micro-lens arrays with varying focal lengths \cite{yanny2020miniscope3d}, local color sensing \cite{arguello2021shift}, metasurface-optics design  \cite{tseng2021neural} and conjugate adaptive optics \cite{mertz2015field}.

In this work, we propose and implement a differentiable propagation model providing a platform to optimize diffraction optical systems with unprecedented design freedom using backpropagation. For example, we perform PSF engineering for thin plate elements placed in any plane in an optical system, design spatially-variant PSFs, perform cascaded optimization of multiple planes and optimize multiple element-placement jointly with their phase functions.  These capabilities are all made possible by implementing a differentiable version of the Collins integral using Pytorch \cite{paszke2019pytorch}. This approach, while still within the boundaries of the thin-element and paraxial approximations, offers several advantages: (1) It is computationally efficient, which is made possible by alleviating strict sampling requirements of Fourier optics and requiring a few FFT operations, (2) it allows the optimization of the positions and functions (\emph{i.e} phase/amplitude profiles) of many elements in a cascaded optical system comprising of lenses and other thin elements, (3) it enables the optimization of diffraction limited systems and offers PSF design freedom that cannot be achieved using geometrical optics, and (4) a hybrid approach can be readily superimposed on our method, where thick-elements and spatial aberrations (\emph{e.g.} from a ray tracing software) can be accounted for. We apply our method numerically and experimentally to several design challenges focusing on shift-variant PSF engineering in microscopy and machine vision.

\section{Cascaded diffraction model} \label{sec:model}

Computing free space propagation under the Fresnel approximation requires evaluation of the Fresnel integral \cite{goodman2005introduction} in each transition between elements:

\begin{multline}
    U_{out}\left(r_{2}\right) = \frac{\exp\left(i k\cdot z\right)}{i\lambda\cdot z}  \cdot \exp\left(\frac{ik}{2z}\cdot\left|r_{2}\right|^{2}\right) \cdot  \int_{P_{1}} U_{in}\left(r_{1}\right) \cdot\\
    \exp\left(\frac{ik}{2z}\cdot\left|r_{1}\right|^{2}\right) \cdot
    \exp\left(\frac{-ik}{z}\left(r_{2}\cdot r_{1}\right)\right)dr_{1},
    \label{eq:Fresnel}
\end{multline}
where  \(U_{out}\left(r_{2}\right)\)  is the scalar electrical field in the output plane \(r_{2}=\left(x_{2},y_{2}\right)\), \(U_{in}\left(r_{1}\right)\) is the scalar electrical field in the input plane \(r_{1}=\left(x_{1},y_{1}\right)\) with spatial extent of \(P_{1}\). \(k=\frac{2\pi}{\lambda}\) is the wavenumber and \(z\) is the paraxial propagation distance.

Numerically evaluating \cref{eq:Fresnel} is challenging, mainly because of the chirp functions (\(\propto\exp{\left(i\cdot x^2\right)}\)) which are not band-limited and the Nyquist frequency which increases with the aperture, \emph{i.e.} larger apertures require finer sampling. Efficient methods to evaluate \cref{eq:Fresnel} have been extensively studied \cite{zhang2020analysis}. In this work we focus mainly on an angular-spectrum method termed the Collins integral \cite{collins1970lens}. This formulation relates geometrical optics ray tracing to the Huygens-Fresnel wave propagation theory. Specifically, we use a scaled version, similar to the approach by G. A. Tyler and D. L. Fried \cite{tyler1982wave}, which was developed to handle atmospheric turbulence computations, and allows for some flexibility in choosing the sampling frequency and grid size.

\subsection{Fresnel propagation in a complex optical system} \label{subsec:Collins}
At the heart of our method is the calculation of diffraction between planes of interest (\emph{i.e.} apertures, phase or amplitude elements) in an optical system. We now derive an expression for the propagation in segment \textit{j} between two planes in the system. In geometrical optics, the well-known ABCD matrix \cite{brouwer1964matrix} defines the transformation of a light ray (defined by lateral position and angle) between two planes. We write the ABCD matrix as a 2X2 matrix, assuming rotational symmetry: 
\begin{equation}
    \mathcal{M}_{j} = 
    \begin{pmatrix}
    A_{j} & B_{j} \\ C_{j} & D_{j}
    \end{pmatrix}
    .
\end{equation}
Using the ABCD matrix, the Collins integral defines the wave propagation as follows \cite{collins1970lens}: 

\begin{multline}
    U_{out}\left(r_{2}\right) = \frac{\exp\left(ik\cdot z\right)}{i\lambda\cdot B_{j}} \cdot \int_{P_{1}} U_{in}\left(r_{1}\right)  \cdot \\
    \exp\left(\frac{ik}{2B_{j}}\cdot\left(D_{j}\cdot \left|r_{2}\right|^{2}-2r_{1}\cdot r_{2}+A_{j}\left|r_{1}\right|^{2}\right)\right)dr_{1}.
    \label{eq:Collins_org}
\end{multline}
A scaling parameter \(m\) can be introduced by defining a scaled grid \( r_{2}'=r_{2}/m \). Plugging this into \cref{eq:Collins_org} gives the scaled Collins integral which is the main tool in this work (see supplementary  section A.1 for the full derivation):

\begin{multline}
    U_{out}\left(r_{2}\right) = 
    \frac{\exp\left(\frac{ik}{2\cdot B_{j}}\cdot(D_{j}-\frac{1}{m}) \cdot\left|r_{2}\right|^{2}\right)}
    {i\lambda\cdot B_{j}} \cdot \\ \int_{P_{1}} U_{in}\left(r_{1}\right) \cdot 
    \exp\left(\frac{ik}{2\cdot B_{j}}\cdot(A_{j}-m) \cdot\left|r_{1}\right|^{2}\right)
    \cdot \\
    \exp\left(\frac{i \cdot k\cdot m }{2 B_{j}}\left|\frac{r_{2}}{m} - r_{1}\right|^{2}\right)dr_{1}.
    \label{eq:Collins}
\end{multline}
 The integral is defined for sections where \( B_j \neq 0 \). The numerical evaluation of \cref{eq:Collins} can be simply computed \emph{via} a Fourier transform \cite{schmidt2010numerical}, where the input and output planes are sampled by intervals \( \delta_{1} \) and \( \delta_{2} \), respectively, and \(m=\frac{\delta_{1}}{\delta_{2}}\):

\begin{flalign}
\begin{split}
    & U_{out}\left(r_{2}'\right) \propto Ch_{out} \cdot \mathcal{F}^{-1} \left[\mathcal{F}\left( U_{in}\left(r_{1}\right) \cdot Ch_{in} \right) \cdot Ch_{tf}\right]; \\ 
     \\ 
    & Ch_{in} = 
    \exp\left(\frac{ik}{2\cdot B_{j}}\cdot(A_{j}-m) \cdot\left|r_{1}\right|^{2}\right),
    \\
    & Ch_{tf} = 
    \exp\left(\frac{-i\pi\lambda B_{j}}{m}\cdot\left|f\right|^{2}\right),\\
    & Ch_{out} = 
    \exp\left(\frac{ik}{2\cdot B_{j}}\cdot(D_{j}-\frac{1}{m}) \cdot\left|r_{2}'\right|^{2}\right).
    \label{eq:Collins-FFT}
\end{split}
\end{flalign}
where \(f=\left(f_x,f_y\right)\) is the sampling grid in the Fourier domain. To produce a differentiable diffraction model from an input plane to an off-axis position on an output plane, we use the matrix \( \mathcal{M}_{j} \) to shift the coordinate system \(r_{2}\) \cite{gross2020cascaded} by tracing the geometrical chief ray. Then we evaluate \cref{eq:Collins} to get the diffraction pattern. A minor projection correction (for small angles) can be considered here, we neglect this effect to conserve memory. We note that the Collins integral can alternatively be evaluated as a real-space convolution with fixed sampling frequencies. 
Finally, if the output plane is not the final image plane, an additional step of multiplying with the shifted element at the output plane and re-centering is performed to enable the cascaded propagation to the following sections in the system.

Another important question to address is how to simulate the point source. Modelling the input source correctly is essential in creating an accurate propagation simulator. One simple choice would be to use a 2D Gaussian at the input plane; however, this is an unsatisfactory choice as the Fourier transform will lack the high frequency content of a point source. We use two alternatives approaches, detailed in supplementary section A.2, the first is to model the electrical field of a paraxial point source at the first lens, and the second is to backpropagate the electrical field from a defocused aperture, which enables simple introduction of system aberrations. 

\subsection{Comparison with other propagation methods}
\label{subsec:whyCollins}

The key advantage of using the Collins integral is that it enables skipping entire lens stacks to calculate the diffraction between thin-planes of interest - all it requires is the relatively simple-to-obtain ABCD matrix, while most other methods require expensive computations between each element. To further justify using Collins-integral over other common methods, we examine several numerical issues. First, in principle, Fresnel diffraction can be evaluated as a convolution integral by directly solving \cref{eq:Fresnel} in a single step. However, in cascaded systems, this imposes two main challenges: (1) the sampling of the output is fixed to \( \delta_{2} = \frac{\lambda\cdot z}{N\cdot \delta_{1}} \), requiring different zero padding for a set of wavelengths and interpolations to match given restrictions (\emph{e.g.} camera pixel size or desired sampling in a Diffractive Optical Element (DOE)), (2) correct sampling of the chirp function inside the integral imposes a limit on the maximal propagation distance \(z\geqslant \hat{z} = \frac{N \cdot \delta_{1}^2}{\lambda}\); at the same time, in a cascaded scenario (where a correct complex field is important, and not just the amplitude), the outer chirp can also induce aliasing. This creates a severely limiting situation where the calculation is correct only for \(z = \hat{z}\).

Other existing approaches for calculating Fresnel propagation involve a two-step calculation. Solving \cref{eq:Fresnel} as a convolution integral can be done by sampling the transfer function in Fourier space (usually for short propagation distances) or in real space (usually for long propagation distances). One way to mitigate the sampling constraints is by doing two single-step propagations \cite{coy2005choosing,zhang2004algorithm}. All those methods and their variations have different sampling requirements and often requiring drastic zero-padding to reduce aliasing \cite{zhang2020analysis}. When simulating free space propagation these methods can be viable; However, the Collins
approach allows for fewer propagation calculations and choosing planes with minimal field supports to reduce the sampling constraints. 

\begin{figure}[ht!]
\centering
\includegraphics[scale=1.0]{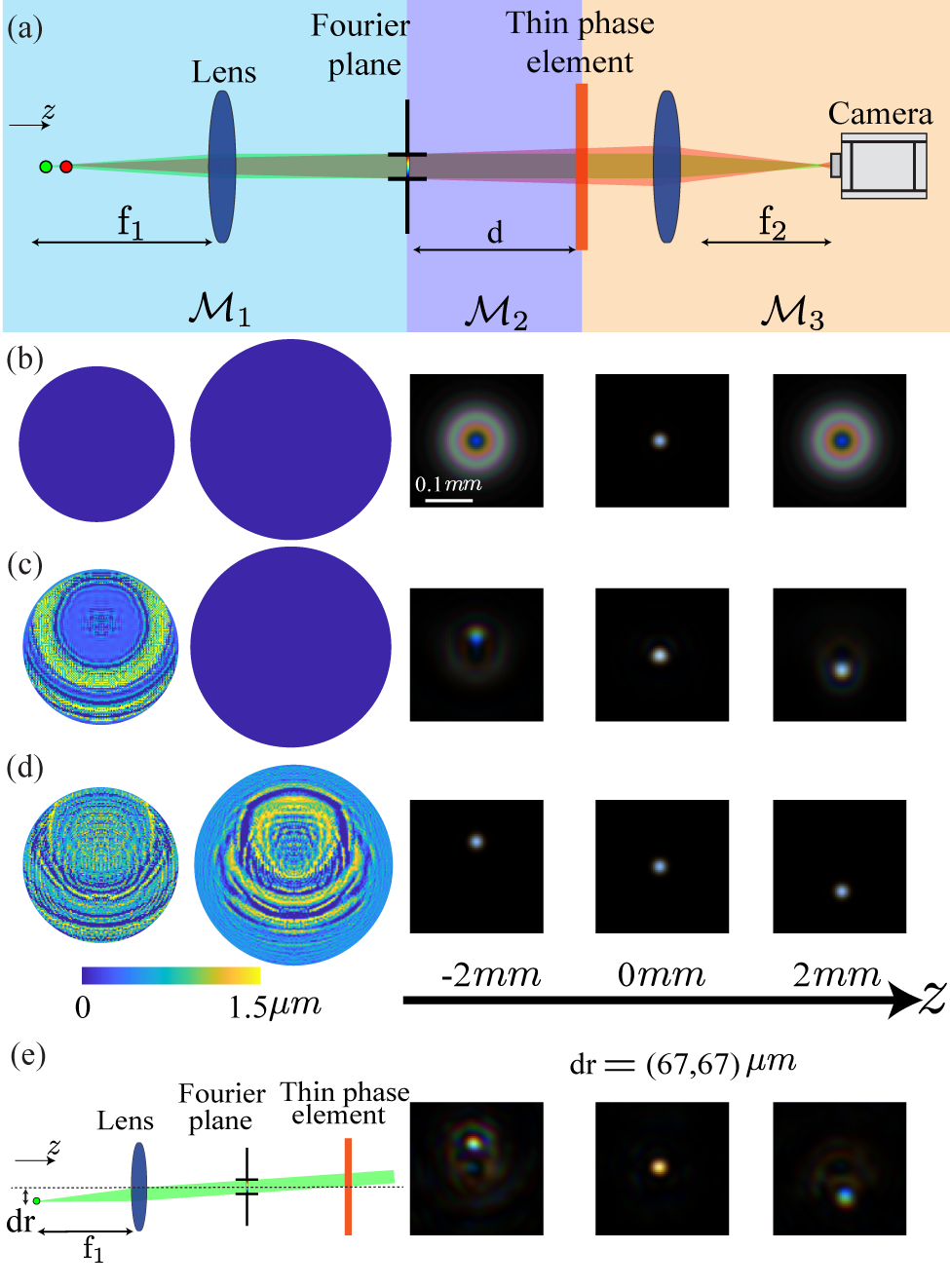}
\caption{Simulation of a 4-f optical system with two thin phase plates, one placed in the back focal plane and the second a distance d away. (a) The system is divided into three parts with corresponding matrix \(\mathcal{M}_{j}\). (b) The RGB PSF of the unmodulated system. (c) The solution of a phase retrieval task \cite{ferdman2020vipr} with the aim to create a multi-focus RGB PSF, where only the mask in the back focal plane is optimized. (d) Solution to the same task but two elements are jointly optimized to correct the color drift.  (e) Induced off-axis aberration which is caused by the second DOE in (d). }
\label{fig:sim1}
\end{figure}
\subsection{Numerical example: multi-focus RGB PSF design}
\label{subsec:MultiFoc}
As a first example, we consider the case of optimizing a multi-focus RGB PSF \emph{via} phase retrieval, simulation presented in \cref{fig:sim1}. The challenge here is to design an optical system that maps a point in 3D object space to a point in 2D image space, where the lateral position in image space depends on the axial object-space position; and, importantly, different colors are to be mapped identically. We start from this challenge \cite{soifer1994multifocal} because it has been shown that a multi-focus PSF can be produced for a single color with a phase mask in the Fourier plane; however, such a system suffers drastic color aberrations for Fluorescent sources with a spectral width of a few tens of nm \cite{abrahamsson2013fast}.

In \cite{abrahamsson2013fast}, a chromatic correction element was placed in conjunction with a binary phase mask to solve this issue. In this section, we show that our model can jointly optimize non-binary phase modulation and chromatic corrections on-axis. A simple 4-f system is presented in \cref{fig:sim1}(a). Three on-axis RGB (discrete wavelengths of 460,550,640 \(nm\)) point sources are considered (in the figure, the two plotted points  indicate change in axial position of a multicolored source). The first lens has a focal length of 5 cm, the second lens has a focal length of 10 cm and the aperture diameter is 2.5 mm, which is a typical for a Diffractive Optical Element (DOE). The unmodulated RGB PSF of this system is plotted in \cref{fig:sim1}(b). The considered DOEs are simulated with maximal height of \(h_{max} = 1.5\) \( \mu m\), having the phase delay function (when placed in air)
\begin{equation}
    \Phi\left(x,y\right) = k \cdot \left(n\left(\lambda\right)-1\right)\cdot h\left(x,y\right),
\end{equation}
where \(n\left(\lambda\right)\) is the refractive index of the DOE material (Fused Quartz in this case). 

 A first DOE \( h_{1} \) is placed at the aperture (Fourier plane) while an optional second DOE \( h_{2} \) is placed a distance \( d = 7.5\) \(cm\) after the aperture. We used \cref{eq:Collins-FFT} to calculate the diffraction in all sections in \cref{fig:sim1}(a), although we note that section \(j=2\) can be simply performed by most other methods. To achieve a multi-focus PSF, we minimize the cost function: 

\begin{multline}
    \mathcal{L}_{\text{PR}}\left(h_{1},h_{2}\right) =  \frac{1}{N_p\cdot N_z\cdot N_c}\sum\limits_{n=1}^{N_p}\sum\limits_{m=1}^{N_z}\sum\limits_{l=1}^{N_{c}}\|\text{PSF}\left(n;h_{1},h_{2},z_{m},C_{l}\right)\\ - \text{PSF}\left(n;h_{1}=0,h_{2}=0,z=0,C_{l}\right)\|_2^2 ,
\end{multline}
where \( n= 1,2...N_p\) are the pixels in the cropped FOV, \( \text{PSF}\left(n;h_{1},h_{2},z_{m},C_{l}\right) = \| U_{out}\left(n;h_{1},h_{2},z_{m},C_{l}\right)\|^{2} \) is the  PSF at the image plane for a point source with defocus \( z_{m} \) and wavelength \(C_{l}\).  \(\text{PSF}\left(n;h_{1}=0,h_{2}=0,z=0,C_{l}\right)\) is the diffraction limited PSF (in-focus) per wavelength \(C_{l}\). 
\begin{figure*}[ht!]
\centering
\includegraphics[scale=1.1]{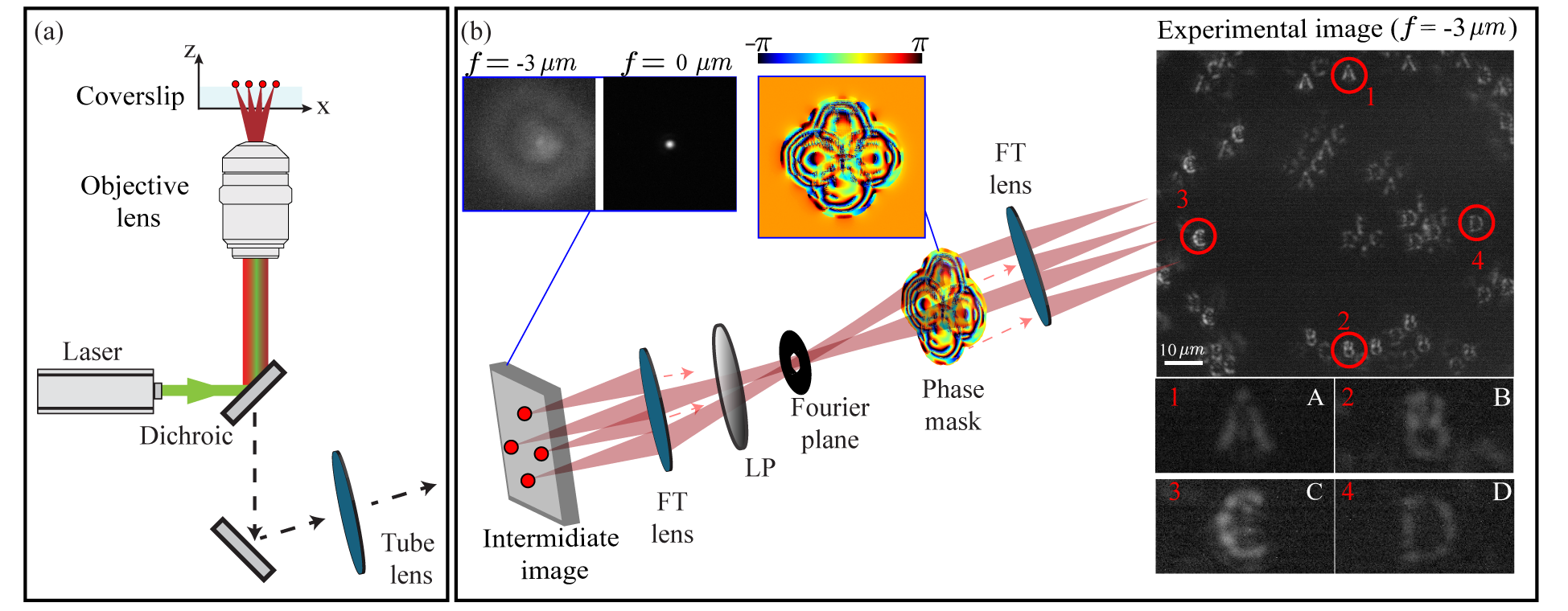}
\caption{Experimental out-of-aperture phase retrieval. (a) A standard inverted microscope with 40X 0.75NA objective. (b) emission from fluorescent beads passes through the microscope and an extra extended 4f system with a LC-SLM placed 9.6 cm behind the back focal plane. We chose four lateral positions (37.5 \(\mu m\) in each direction from the optical axis) in a single defocus plane (-3 \(\mu m\)) to optimize the PSF of a fluorescent molecule placed there. Each position was assigned a letter in the phase retrieval algorithm, resulting in the phase mask plotted. The experimental image was acquired by finding a FOV with beads close to the designed positions. Intermediate lateral positions show a combination of the PSFs, as expected for positions that were not optimized. The contrast of the full FOV image was
adjusted for better visibility.}
\label{fig:exp1}
\end{figure*}
The optimization is performed with a continuous height map for the DOEs, and a softmax function with decreasing temperature is applied to the height map such that at the final iterations the mask will get sampled on a possible fabrication grid with an axial step size of \( 100\) \( nm \). The optimization is performed using the Adam optimizer \cite{kingma2014adam} until the cost function stagnates. 
First, we optimized the Fourier plane alone, resulting in the phase mask and corresponding PSFs in \cref{fig:sim1}(c), which manages to shift and refocus the input field, however a chromatic shift between the colors is noticeable. We note that theoretically increasing \(h_{max}\) can achieve any chromatic correction in a single element; however, DOEs with large height differences are very challenging to fabricate and exhibit large errors. Next, we jointly optimized both elements, resulting in the DOEs and corresponding PSF in \cref{fig:sim1}(d). The entire optimization takes two minutes (on a laptop with Intel i7-8750H and NVIDIA 8Gb 2070RTX GPU). Optimizing two planes enabled the correction of the color shift; however, we only considered on-axis PSF optimization, which results in a significant off-axis aberration, shown in \cref{fig:sim1}(e). This means that the current approach can be used to readily optimize multiple elements in a complex optical system for on-axis propagation, \emph{e.g.} in holography \cite{javidi2021roadmap} or laser mode manipulations \cite{maluenda2013reconfigurable}, however, in this section we sampled the PSF only on-axis. This is not a suitable approach for imaging systems, which are the focus of the rest of this paper. The intermediate conclusion is that addressing and optimizing the shift-variant PSF (by sampling PSFs at different lateral positions) is crucial for making this method viable, thus, this is the approach we take in the remainder of this work.

\subsection{Numerical considerations}
\label{subsec:numres}

A main consideration in this work is the computational burden which comes from FFT computations of large matrices. As we consider spatially variant PSFs, our matrices have five dimensions \( \left(N_{xy},N_{z},N_{c},H,W \right) \) being the lateral position, axial position, wavelength and image height and width, respectively. The size of such matrices can be limited by memory constraints, and thus we chose our sampling grids to minimize \(\left(H,W\right)\). Full description of the sampling conditions is provided in supplementary section D.

\begin{figure}[ht!]
\centering
\includegraphics[scale=1.0]{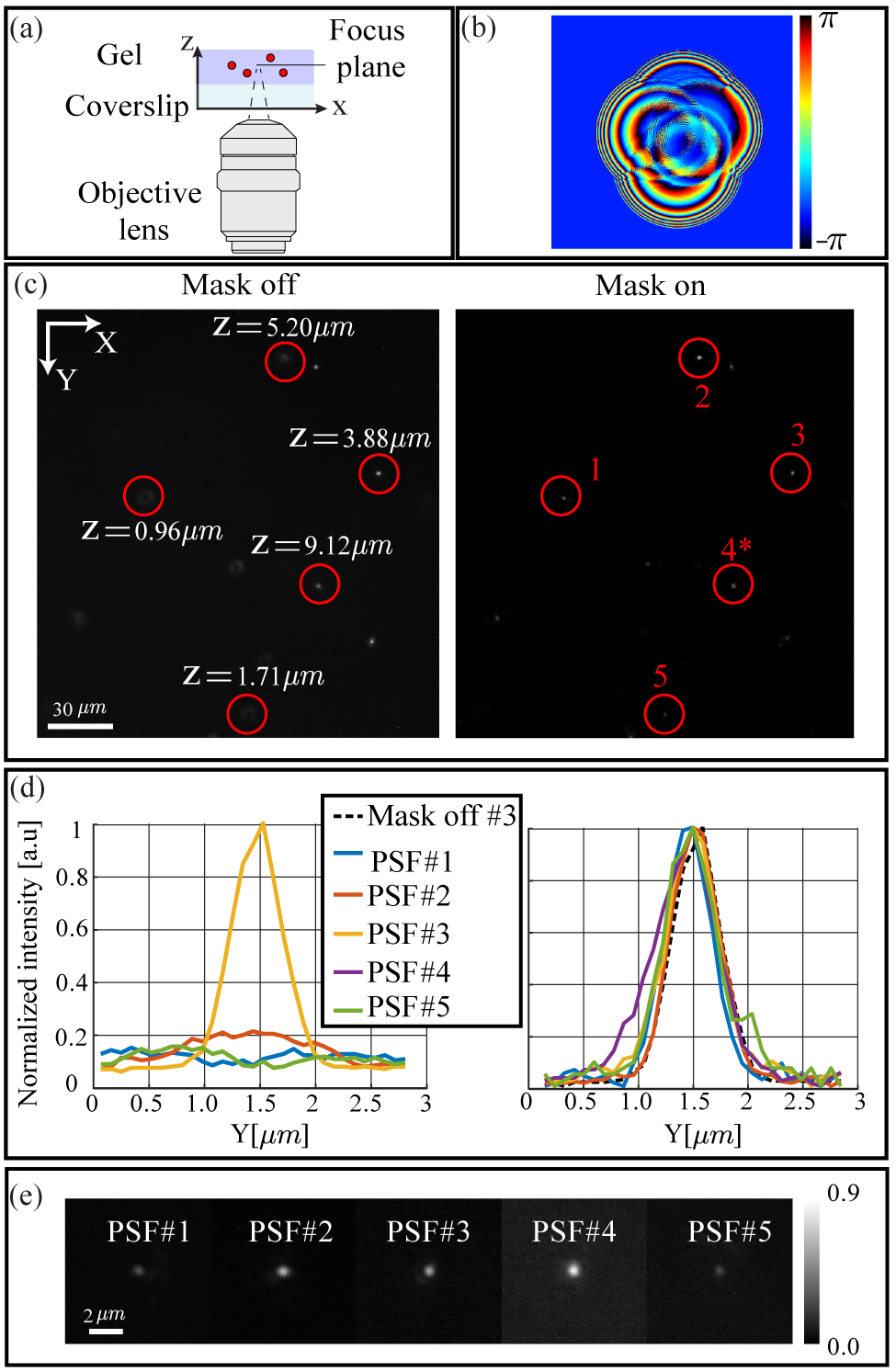}
\caption{Refocusing experiment. (a) Beads in Gel creating a sparse 3D sample. The optical setup is similar to the one in \cref{fig:exp1}(a-b). (b) The phase retrieved mask which was applied to the image in (c-left) and refocused to (c-right, contrast adjusted for visibility). In both images, the objective lens is focused close to the axial plane of emitter \(\#3\) (at z = 3.7 \(\mu\) m). Note especially beads \(\#1,2,5\) that significantly improve visibility after refocusing. Emitter \(\#4\) in the refocused image is not the emitter in the circle on the left, but a defocused one which cannot be seen on the left image - both emitters have a very similar lateral position, which can be seen the full z-stack supplementary Fig. 5.(d-left) Line plots if the emitters in (c-left) without emitter \(\#4\) ) and (c-right) line plots of the refocused PSFs. (e) Cropped PSFs from (c), normalized relative to the z-stack focused image (per emitter)}
\label{fig:exp2}
\end{figure}

As chirp functions are not band limited, the main constraint comes from avoiding aliasing inside the limiting apertures at the input and output planes. We note that this consideration needs to be very carefully examined for each sub-system separately, as aliasing can occur if the input field adds high enough frequency at the edges of the apertures, \emph{i.e.} reducing the computational complexity comes at the cost that each diffraction calculation will be correct only inside the apertures and \emph{w.r.t} input fields with a pre-defined bandwidth ( limits the axial range and possible aberrations). For example, in the simulation provided in \cref{fig:sim1} we used \((H,W) = (512,512)\) sampling points, but a larger defocus in the input plane or using high frequency (\emph{e.g.} gratings) components could necessitate a drastic increase in \(\left(H,W\right)\). 

\section{Experimental demonstrations}
\label{sec:exp}
In this section we detail two experiments performed to validate the capabilities of the method to design thin elements, focusing on phase masks. We used an epi-fluorescence microscope (40X magnification and 0.75 NA) with an extended 4f (two achromatic lenses with focal lengths of 10 and 20 cm) system as shown in  \cref{fig:exp1}(a). A Liquid-Crystal Spatial Light Modulator (LC-SLM, Pluto VIS-130 Holoeye) is placed 19.6 cm after the first lens, namely, downstream from the Fourier plane, behind an aperture (to filter stray light) matching the back focal plane size of the objective lens (3.75 mm). Our calibration process consists of two steps, the first is calibrating the microscope aberration which is based on our previous work \cite{ferdman2020vipr}. The second step is registration of the theoretical and experimental ray tracing of the chief ray from the object plane to the LC-SLM plane, for further details see supplementary section E.4.

\label{sec:exp-test}

\subsection{Out-of-aperture phase retrieval} \label{subsec:Letters} 
The first experiment validates our out-of-aperture phase retrieval method by designing spatially variant PSFs at a specific defocus position. We use Flourescent beads (Introvigen, 625/645 200 nm) adhered to a glass coverslip as our sparse point source sample (see supplementary section E.2 for details about the sample preparation). The cost function was altered to optimize a single color (matching the beads' peak emission of \( 645\) \( nm \)) phase mask \( \Phi\) instead of DOE heights. 
\begin{multline}
    \mathcal{L}_{\text{PR}}\left(\Phi\right) = \\ \frac{1}{N_p\cdot N_{xy}}\sum\limits_{n=1}^{N_p}\sum\limits_{q=1}^{N_{xy}}{\|\text{PSF}\left(n;\Phi,xy_{q},z\right) - L\left( n;xy_{q}\right)\|_2^2} \\
    + \alpha \cdot TV\left(\Phi\right).
\end{multline}
The second term in the optimization is an anisotropic Total Variation (TV) loss on the phase mask, to enforce smoothness which improves the performance in the case of using LC-SLMs \cite{moser2019model}. The cost function term \( L\left( n;xy_{q}\right) \) was designed by 
choosing four lateral positions \(xy_{q}\) at a defocus \(z\) in the FOV (limited by the LC-SLM) and assigning a picture of a letter as the desired PSF. The results can be seen in \cref{fig:exp1} where emitters close to the designed positions exhibit the desired PSF, while intermediate positions exhibit PSFs that are a shifted combination of the letters (as they sample intermediate positions in the phase mask plane). Furthermore, we repeated the experiment with eight letters spanned over four lateral and two axial positions. The results are presented in supplementary section C, showing that this method can design shift-variant 3D PSFs; however, aberrations were more prevalent when simultaneously optimizing eight letters resulting in lower visual quality.

\subsection{Analog refocusing} \label{subsec:Gel}

The second experiment validates the estimation of positions and defocus correction by spatially refocusing single emitters in a sparse 3D object space. The challenge here is how to produce a single 'in-focus' image of several objects in the field-of-view which are at different axial positions. Specifically, in our case, the depth of field of the system is \( \approx 1.5\) \(\mu m \) and we were able to image objects spanning a 9 \(\mu m\) axial range. We used a sparse bead sample embedded in Gel (see supplementary section E.2 for sample preparation details) to create a 3D object. First, a focal stack was acquired by shifting the objective lens over the axial range. The focal stack was used to estimate the 3D positions of the emitters (see supplementary section E.3). Then, we chose some of the emitters spanning different z position ( \emph{e.g.} 5 in the case presented in \cref{fig:exp2}) and optimized a phase mask where the cost function was aimed at correcting the defocus phase per emitter ,\emph{i.e.} compare the theoretical defocused PSF to the in-focus PSF at \( z= \text{AFP}\) where AFP is the actual focus plane for an objective focused at \( 3.7\) \( \mu m \) above the coverslip (accounting for the refractive index mismatch between the Gel and air): 

\begin{multline}
    \mathcal{L}_{\text{PR}}\left(\Phi\right) =  \frac{1}{N_p\cdot N_{xy}} \sum\limits_{n=1}^{N_p}\sum\limits_{q=1}^{N_{xy}}\|\text{PSF}\left(n;\Phi,xy_{q},z_{q}\right) - \\ \text{PSF}\left( n;\Phi=0,xy = 0,z=AFP\right)\|_2^2
    + \alpha \cdot TV\left(\Phi\right).
\end{multline}
In this case we chose to bin the camera pixels (to an effective size of \(6.9 \mu m\)) to increase the signal in the ground truth estimation and to reduce computation time. After the acquisition of the ground-truth positions, the optimization ran for two minutes and the phase mask was then immediately applied, so that drift can be minimized.  As the emitters beams overlap in the phase mask plane, some contrast is lost due to induced aberrations. Comparing to the in-focus emitter \(\#3\), it lost 30\% of the max intensity. The intensity loss ranges 10-65\%, depending on the acquired aberration. We chose to use our approach even though this specific case could have been solved semi-analytically by calculating the appropriate defocus phases from \cref{eq:Collins} and super-imposing them (in the correct lateral position) on the LC-SLM. The results from this section show the applicability of our method for designing Micro Lens Arrays (MLA) with different focal planes and spatial positions \cite{yanny2020miniscope3d,zhang2021dilfm} and dynamic analog refocus capabilities for sparse samples which surpass a single Extended Depth Of Field (EDOF) phase mask \cite{nehme2021learning} for sparse samples in terms of photon efficiency, axial range and aberration corrections (which can be applied to each emitter separately).  

\section{End-to-end design}
\label{sec:end-to-end}

In this section we use our approach to perform end-to-end learning in a computational imaging task. As a benchmark, we apply our model to the task of semantic segmentation on the Berkeley deep drive database (BDD100K) \cite{yu2020bdd100k}. The database consists of 100K images of size 720x1280 pixels in 3 color channels, captured by a camera located on the dashboard of a driving car. Each image in the database is segmented to 19 different classes, such as: road, building, car, bicycle, etc. Previous work by E. Tseng A. Mosleh, and F. Mannan 2021 et al. \cite{tseng2021differentiable} showed that optimizing a lens-stack with end-to-end learned geometrical optics design which considers the spatial PSF can improve the results in object detection tasks in computer vision. Our logic behind choosing this application is that using a shift-variant PSF based on diffractive optics could potentially exploit the typical structure in such scenes (the sky is at the top of the image, the road is on the bottom, etc.) to better encode the class of each region in the image. Considering the structure in such scenes, it is likely that adding the additional degree of freedom of allowing the PSF to vary throughout the field-of-view would be advantageous. 

We model image acquisition as an incoherent system. Calculating the spatially variant PSF per-pixel and performing convolution with a full frame is computationally infeasible, thus we adopt a linear PSF interpolation approach \cite{hirsch2010efficient}. Under this model, the PSFs are calculated on a grid and then convolved with a weighted image, and then summed to generate the modulated image \(I_{mod}\left(n,m;C_{l}\right)\), \emph{i.e.}:
\begin{multline}
    I_{mod}\left(n,m;C_{l}\right) = \sum\limits_{k=1}^{K} \text{PSF}_{k}\left(n,m;h,z=0,C_{l}\right)*\\
    \left( \omega_{k}\left(n,m\right) \cdot I_{0}\left(n,m;C_{l}\right)\right), 
    \label{eq:lin_weight}
\end{multline}
where \(*\) is a 2D convolution, \(C_{l}\) is the color channel, \(\omega_{k}\left(n,m\right)\) is the window weighting function associated with \(\text{PSF}_{k}\) which is the spatially variant PSF calculated at a chosen grid point and \(I_{0}\left(n,m;C_{l}\right)\) is the input image at pixel \(\left(n,m\right)\) and wavelength \(C_{l}\). The amount of chosen kernels (\(K\)) creates a trade-off between computational efficiency and accuracy. The faster the PSF changes (spatially), the denser the kernels need to be sampled to accurately represent it. In other methods, a SVD approach can be taken to estimate the amount of kernels \cite{yanny2020miniscope3d} needed, or an optimized weighting \cite{denis2015fast} grid based on the PSF can be computed. We chose a rather sparse kernel grid (K=24 for each color) and linear weighting to enable reduced memory usage and allow faster computation. After the initial system design, accuracy can then be improved by retraining the net with calibrated or densely sampled PSFs. An example phase retrieval for a point grid is presented in supplementary section A.3. The position of the mask defines the chirp function phase, and as such we can compute the gradient of the position \emph{w.r.t} the PSF. Thus, importantly, we allow the network to learn not only the phase mask, but also its position, which is a degree of freedom rarely explored in the realm of Fourier optics. 

Our end-to-end model is composed of two main blocks: the optical block and 24-layer ResNet (see supplementary fig. S2). The optical block depicts the formation of the final image in the camera as described above, then, the image is passed through the ResNet network that generates pixel-wise class predictions.
Our loss function consists of three terms:

\begin{equation}
    \mathcal{L}_{\text{net}} = \mathcal{L}_{\text{foc}} +
    \alpha_{TV}\cdot\mathcal{L}_{\text{TV}} + \alpha_{PSF}\mathcal{L}_{\text{PSF}},
\end{equation}
where \(\mathcal{L}_{\text{foc}}\) is a focal loss  \cite{lin2017focal} defined as:
\begin{equation}
    \mathcal{L}_{\text{foc}} = -\alpha_t\cdot(1-p_t)^\gamma \cdot log(p_t),
\end{equation}
where \(p_t\) is the predicted probability of the pixel to have the class t, \(\gamma\) reduces the loss for well classified classes and \(\alpha_t\) controls the weight of each class. The focal loss is similar to the Cross Entropy loss,  while trying to cope with the problem of class imbalance in the database.

\(\mathcal{L}_{\text{TV}}\) is a Total Variation loss on the network output and  \(\mathcal{L}_{\text{PSF}}\) is a regularization on the PSF defined as: 

\begin{equation}
    \mathcal{L}_{\text{PSF}} = \sum\limits_{l=1}^{N_c}\sum\limits_{n=R}^{H}\sum\limits_{m=R}^{W} \text{PSF}\left(n,m;h,z=0,C_{l}\right),
\end{equation}
where R is the cropping region for the PSF chosen to be 225 pixels in this case. We chose to reduce the size of the PSF to reduce high frequencies in the phase mask and reduce the memory usage in the computation of \cref{eq:lin_weight}.

\begin{figure*}[ht!]
\centering
\includegraphics[scale=0.9]{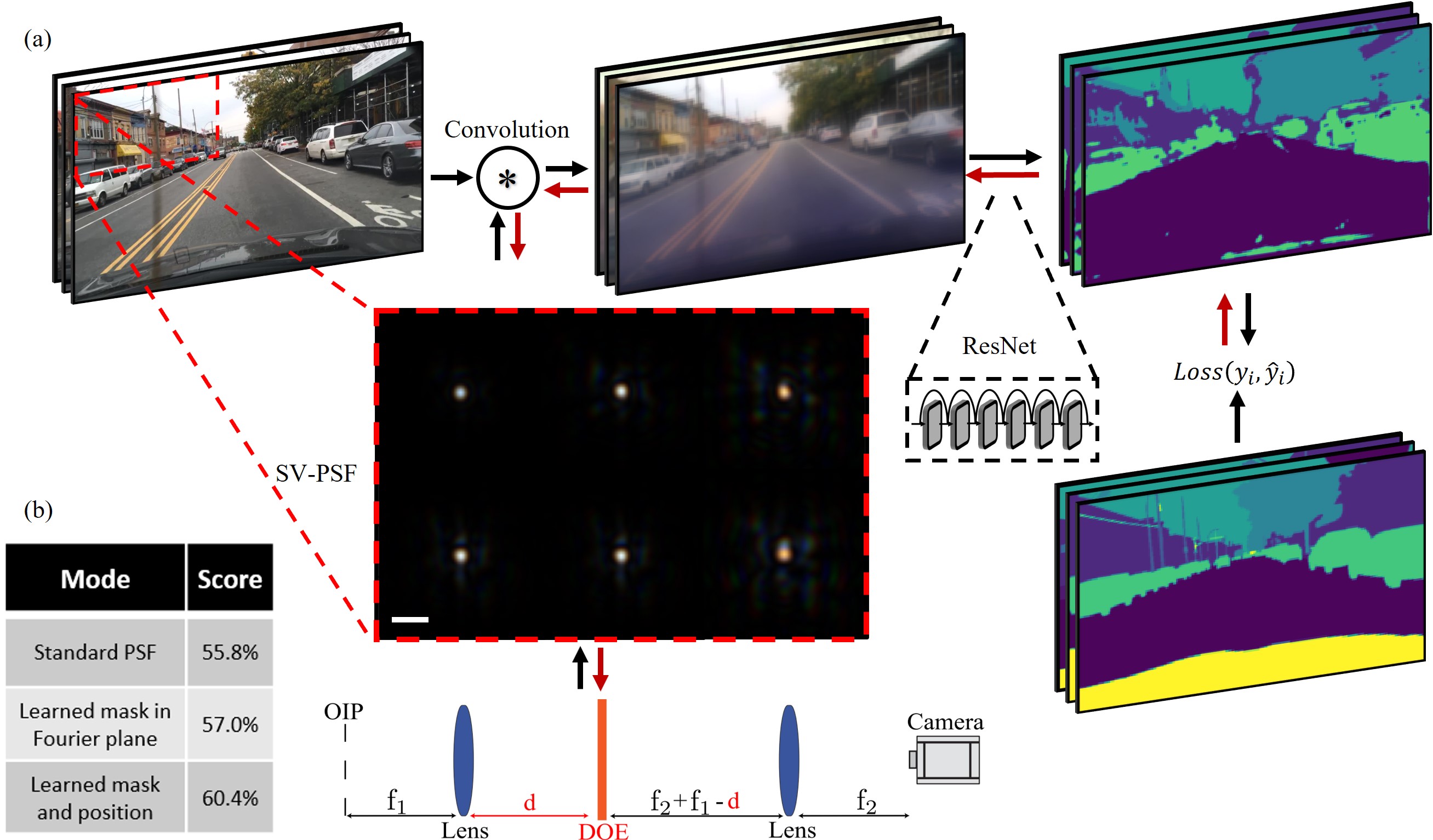}
\caption{End-to-end learning for semantic segmentation. (a) In the forward pass, the input image is convolved with the PSF (24 kernels per wavelength, we show only 6, gamma corrected for visibility) that was generated by the learnable phase mask in the optical system, where the original image plane (OIP) is the location of the image from the dataset. We learn both the phase mask pattern, and its position (\(d\)) in the optical path. The observed image is inserted to the ResNet network, that generates a pixel-wise class prediction map. Finally, we calculate the loss function between the prediction and the ground truth. In the backpropagation step we update both the ResNet weights and the phase mask parameters. Black/red arrows mark forward/backward paths. (b) The final scores for three different training modes: standard PSF, learned PSF in the Fourier plane, and learned PSF and position. Scale bar is 150 \(\mu m\) in the camera plane}
\label{fig:E2E}
\end{figure*}

To test our network we have compared the performance of 3 different architectures: ResNet with standard imaging conditions, namely, blurring the original image; ResNet with learnable phase mask in the Fourier plane (yielding a shift-invariant PSF); and ResNet with learnable phase mask position and function. In \cref{fig:E2E} we present the experimental scheme: First, we convolve the input image with the system's PSF (either learned PSF or standard PSF). The output of this optical block is the observed image, which is then passed to the neural network that predicts the pixel-wise segmentation of the image. 

For simplicity, our model assumes augmentation of a standard imaging system with an additional 2 lens system, with our phase mask placed at some plane (with axial position optimized by the net) in between (\cref{fig:E2E}(a)). We report the average accuracy that we achieved on 1000 test images by using each architecture (\cref{fig:E2E}(b)). The standard PSF architecture predicted 55.8\% of the pixels correctly. When the phase mask was learned but the position of the mask was constant (in the Fourier plane, \emph{i.e.} a spatially-invariant PSF) the percentage of correctly classified pixels was 57.0\%, namely, achieving some improvement over not learning a phase mask at all. Finally, in the spatially-variant design, the network correctly classified 60.4\% of the pixels - an improvement of 4.7\% (8.2\% relative improvement) compared to standard imaging. The phase mask position was optimized by the net to be ~5.1 cm after the first lens in the 4f system (where the focal distances of the first and second lenses are 15 cm and 5 cm, respectively). Thus, we conclude that a shift-variant PSF that can exploit the structure of the data outperforms both standard imaging and shift-invariant PSF engineering. In practice, such a system will require the fabrication of a larger DOE (approximately 8.5 mm in the long axis compared to a circle with a diameter of 2.5 mm in the aperture), which might increase the costs of fabrication when considering lithography, although recently developed techniques show promising potential for simplified fabrication of large masks  \cite{orange20213d,fu2021etch}.  

The results of this experiment show that, for the objective of semantic segmentation in the BDD dataset, simply adding a learnable phase mask in the Fourier plane is not optimal for information encoding. On the other hand, learning the phase mask position and function jointly, enabling the PSF grid to encode information differently in different parts of the image, improved the neural network prediction of the pixel-wise classes over standard imaging and PSF engineering in the Fourier plane. We note that this specific scenario (effectively 2D imaging and 3 color channels) could have been addressed without an optical block, by increasing the number of parameters in the neural network, and having the network perform digitally the physical operation of the optical setup. Nevertheless, with the optical block adding 250,000 learnable parameters (pixels in the phase mask) to a ResNet with approximately 20 million parameters, we can see an efficient improvement in the results. Interestingly, the output of the optical block can serve as an interpretable intermediate result - or alternatively, it can guide the algorithmic design process, by suggesting an initial image-processing operation to improve the reconstruction net of images obtained by a standard imaging system.

\section{Discussion}

In this work we demonstrated a flexible and differentiable method to design complex optical systems containing diffractive elements. Our approach allows for simple and direct wave-optics propagation and optimization of multiple elements in a cascaded system, using backpropagation. We validated the applicability of our method to perform out-of-aperture phase retrieval for incoherent sources, enabling the design of spatially-variant PSFs with substantial freedom, \emph{e.g.} optimization of the element placement, and demonstrated a potential application for autonomous vehicles, performing end-to-end optimization of a computational imaging system for object segmentation.  

Two computational simulations were demonstrated; first, optimizing a cascaded system with two elements, allowing for a simple optimization of a multi-focus RGB PSF (\cref{sec:model}). Second, we showed  end-to-end design of a computational imaging system for an object segmentation task, where the position of a phase mask is learned in conjunction with the spatially variant PSF (\cref{sec:end-to-end}). In \cref{sec:exp}, we validated our method by designing a phase mask for two different scenarios: designing alphabetical-letter PSFs to demonstrate the pixel-wise optimization capabilities and design freedom, and spatially-variant refocusing, suggesting a possibly useful application involving multi-focus imaging. Our method is based on using paraxial ray-tracing to define the wave propagation \emph{via} the Huygens-Fresnel diffraction. This framework allows computation of cascaded diffraction with spatial information, while supporting fast computations and flexibility in sampling rate, enabling the optimization of complex optical systems. Possible future applications can include online correction of spatially-variant PSFs (\emph{e.g.} spatial adaptive optics), end-to-end learning for tasks with structured data, source separation \cite{metzler2021deep}, coherent system design, and more. In the future, our approach can be combined with geometrical ray tracing, which would allow the consideration of thick-element aberrations, fabrication imperfections, chromatic aberrations and other constraints which are straightforward to include in ray tracing.  Ultimately, these two approaches can be combined into a hybrid model, exhibiting the best of both worlds. 

\section*{Funding}
This project has received funding from the European Union’s Horizon 2020 research and innovation program under grant agreement No. 802567 -ERC- Five-Dimensional Localization Microscopy for Sub-Cellular Dynamics, and is supported by the Israel Science Foundation (grant No. 450/18). Y.S. acknowledges the support of the Zuckerman STEM Leadership Program
\section*{Disclosures} The authors declare no conflicts of interest.
\section*{Supplementary materials}
See Supplementary information for supporting content.
\newpage

\appendix
\title{\Huge{Supplementary information}}
\maketitle

\renewcommand{\thefigure}{S\arabic{figure}}
\renewcommand{\theequation}{S\arabic{equation}}
\setcounter{figure}{0}
\setcounter{equation}{0}

\section{Optical simulation model} \label{SIsec:model}

\subsection{Propagation model}
A common way to perform ray tracing is to define an ABCD matrix which 
transform a light ray \(\left(x,\theta\right)\) between two planes. We write the ABCD matrix as a 2X2 matrix, which assumes azimuthal symmetry in ray propagation in the lateral directions. More generally, the ABCD matrix can be written as a 4X4. For a given section of a system (in air) with a geometrical optics matrix:
\begin{equation}
    \begin{bmatrix}
           x_{2} \\
           \theta_{2} \\
    \end{bmatrix}
    = 
    \begin{pmatrix}
    A_{j} & B_{j} \\ C_{j} & D_{j}
    \end{pmatrix}
    \begin{bmatrix}
           x_{1} \\
           \theta_{1} \\
    \end{bmatrix}
    .
\end{equation}
Collins \cite{collins1970lens} presented a method to compute the Huygens-Fresnel integral in such a system: 

\begin{multline}
    U_{out}\left(r_{2}\right) = \frac{\exp\left(ik\cdot z\right)}{i\lambda\cdot B_{j}} \cdot \int_{P_{1}} U_{in}\left(r_{1}\right) \cdot \\   
    \exp\left(\frac{ik}{2B_{j}}\cdot\left(D_{j}\cdot \left|r_{2}\right|^{2}-2r_{1}\cdot r_{2}+A_{j}\left|r_{1}\right|^{2}\right)\right)dr_{1}.
    \label{eq:Collins_org_SI}
\end{multline}
Following the work of Tyler and Fried \cite{tyler1982wave}, a scaling parameter \(m\) can be introduced by defining a scaled grid \( r_{2}'=r_{2}/m \) and doing the following computation:

\begin{multline}
    D_{j}\cdot r_{2}^2-2r_{1}\cdot r_{2}+A_{j}r_{1}^2 = \\
    D_{j}\cdot \left(r_{2}^2+\frac{r_{2}^2}{m\cdot D_{j}}
    -\frac{r_{2}^2}{m\cdot D_{j}}\right)-2r_{1}\cdot r_{2}+A_{j}\cdot \left(r_{1}^2+\frac{m\cdot r_{1}^2}{A_{j}}
    -\frac{m\cdot r_{1}^2}{A_{j}}\right) \\ 
    = m\left(\frac{r_{2}}{m}-r_{1}\right)^2+\left(D_{j}-\frac{1}{m}\right)\cdot r_{2}^2+ \left(A_{j}-m\right)\cdot r_{1}^2. 
    \label{eq:scale_trick}
\end{multline}
Plugging \cref{eq:scale_trick} into \cref{eq:Collins_org_SI} gives the scaled Collins integral which is the main tool in this work:

\begin{multline}
    U_{out}\left(r_{2}\right) = 
    \frac{\exp\left(\frac{ik}{2\cdot B_{j}}\cdot(D_{j}-\frac{1}{m}) \cdot\left|r_{2}\right|^{2}\right)}
    {i\lambda\cdot B_{j}} \cdot \\ \int_{P_{1}} U_{in}\left(r_{1}\right) \cdot 
    \exp\left(\frac{ik}{2\cdot B_{j}}\cdot(A_{j}-m) \cdot\left|r_{1}\right|^{2}\right)
    \cdot \\
    \exp\left(\frac{i \cdot k\cdot m }{2 B_{j}}\left|\frac{r_{2}}{m} - r_{1}\right|^{2}\right)dr_{1}.
    \label{eq:Collins_SI}
\end{multline}
 The integral is defined for sections where \( B_j \neq 0 \) which means it is not suitable to simulate imaging conditions. The numerical evaluation of \cref{eq:Collins_SI} can be simply computed \emph{via} a Fourier transform \cite{schmidt2010numerical}.

We use the matrix \( \mathcal{M}_{j} \) to shift the coordinate system \(r_{2}\) \cite{gross2020cascaded} by ray-tracing the geometrical chief ray. Finally, we evaluate \cref{eq:Collins_SI} to get the diffraction pattern. A minor projection correction can be performed here, \emph{i.e.} transforming the coordinates based on the incidence angles \( \left(\theta_{x} , \theta_{y} \right)\) by projecting the grid at the output plane (scaling the coordinates by \(\left(\cos{\theta_x},\cos{\theta_y}\right)\)).

\begin{figure}[ht!]
\centering
\includegraphics[scale=1.0]{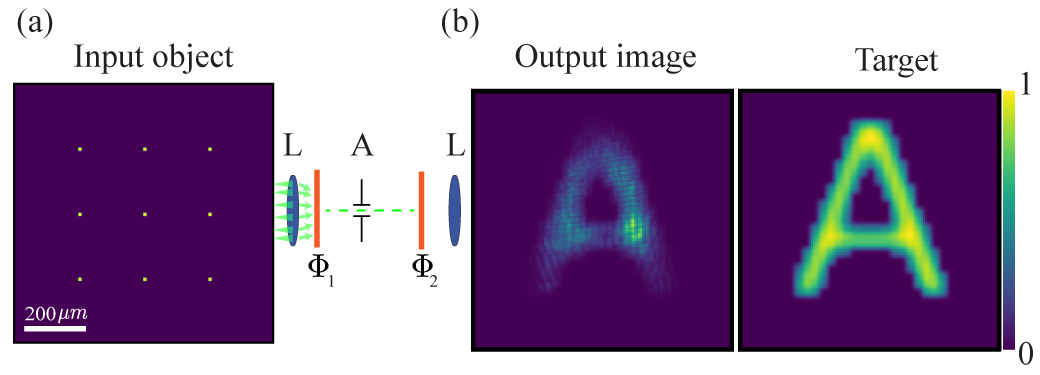}
\caption{Simulation of the image formation model used in the end-to-end design. A grid of 9 points is used as the object, passing through the optical system with 2 phase masks (L denotes lens, \(\Phi\) denotes phase mask and A denotes the aperture). the spatially variant PSF is calculated on those points and a cost is calculated between the obtained image and a blurred letter A.}
\label{fig:SI-gridLet}
\end{figure}

In this work, we neglect this effect to conserve memory by using a uniformly sampled grid in the output plane for all incidence angles.
Finally, an additional step of multiplying with the shifted element at the output plane and re-centering is performed to enable the cascaded propagation to the following sections in the system. The shifting operation can be performed with sub-pixel precision \emph{via} FFT or by simply using a shifted cropping operation (faster but with a pixel resolution), both are differentiable operations and can be used to back-propagate the gradients. 

\begin{figure*}[ht!]
\centering
\includegraphics[scale=1.0]{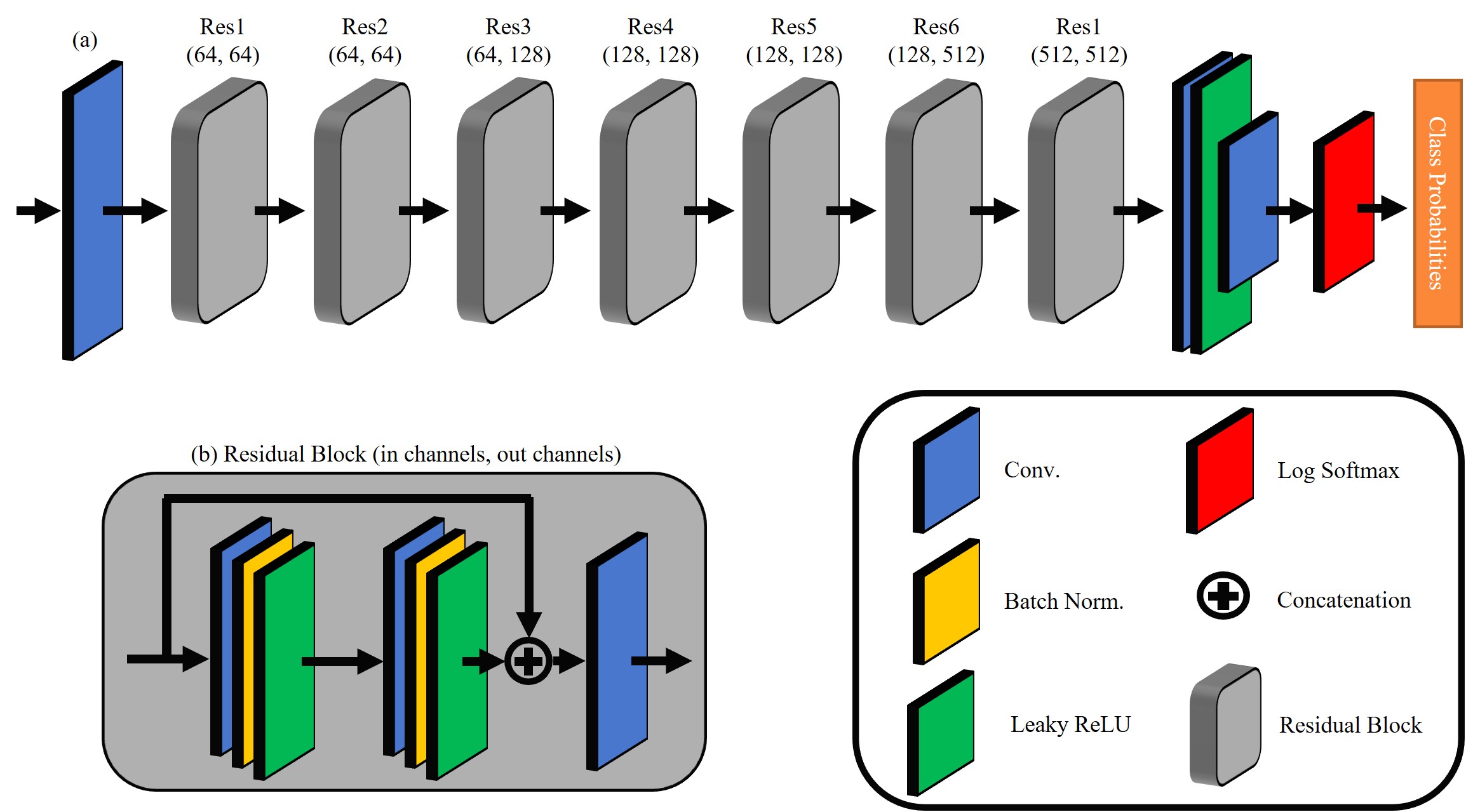}
\caption{Neural network architecture. (a) The first conv. layer has relatively large kernel size to capture features in different scales. Then the data goes through 7 residual blocks. The output of the residual blocks is passed through few more conv. layers and a log-softmax layer to generate the class probabilities per pixel. (b) the architecture of a single residual block. The in channels, mid channels and out channels determine the input and output channels of each conv. layer in the residual block. The conv. layers are separated by batch normalization and leaky ReLU layers.}
\label{fig:SI-NetArc}
\end{figure*}

\subsection{Modelling the input point source} \label{subsec:point-source}

Modelling the input source correctly is essential in creating an accurate propagation simulator. One simple choice would be using a 2D Gaussian at the input plane; however, this is an unsatisfactory choice as the Fourier transform will lack the high frequency content of a point source. 
Under the paraxial approximation, the emission of a point-source is given by a parabloc phase: 
\begin{equation}
        U\left(r\right) = \frac{\exp{\left(i k\cdot z\right)}}{i\lambda \cdot z} 
        \exp{\left(\frac{i k}{2z}r^{2}\right)}.
    \label{eq:paraxial-source}
\end{equation}
As this equation cannot be evaluated at \(z=0\), we implement two options for the input source. For simplicity, we provide the expression for on-axis sources, whereas evaluating off-axis is straightforward by adding a linear phase in the frequency domain. The first approach is by modelling the field, using \cref{eq:paraxial-source} (or an apodized version of it), at the plane of the first element (or lens) in the system, this requires changing \(\mathcal{M}_{1}\) and the sampling conditions. The second approach is starting from a model of the scalar field in the aperture; this is mainly suitable for small defocus values, \emph{i.e.} \(z<f_{1}\) and for telecentric systems which we experimentally focused on. Under this assumption, the Fourier plane scalar field \(E_{\mathcal{F}}\) can be expressed as clear aperture with a phase curvature proportional to the z position: 

\begin{equation}
    E_{\mathcal{F}}\left(r\right) =
    \begin{cases}
    \exp{\left(i k z  \cos\theta\right)},& \text{if } r\leq S\\
    0,& \text{otherwise}
    \end{cases},
    \label{eq:Fourier-field}
\end{equation}
where \(S\) is the limiting aperture and \(\sin{\theta} = r/f_{1}\). For low NA values, this can be further simplified by \(\cos{\theta} = 1-\sin{\theta}^{2}/2\). 

If there is no requirement to propagate the gradient before the Fourier plane (or entrance pupil), then the optimization can directly start from this position, reducing the need the compute the diffraction to that point like in the example provided in main text Fig 1. In other cases, where there exists elements before the Fourier plane or when the Fourier plane is entirely skipped in the calculation, we back-propagated the scalar field to the input plane by noting that \cref{eq:Collins_SI} can be computed in the reverse direction by using the inverse matrix \(\mathcal{M}_{j}^{-1}\).

\subsection{Incoherent image formation}

In this section we show a simple optimization of the full image model with spatially variant PSFs (to show the capabilities of the optical approach without the learning stage) that was used to simulate the results in Main section 4. We chose a similar optical system and optimized two thin phase elements \(\left(\Phi_{1},\Phi_{2}\right)\), placed 1 \(cm\) after the first lens and 1.5 \(cm\) before the second lens. In \cref{fig:SI-gridLet} we show a simple optimization case where the input is a 3X3 point grid. The PSFs are sampled on the grid and a phase retrieval optimization routine (minimizing \(\mathcal{L}_{\text{A}}\) is ran such that the final image (\( I_{mod}\) ) will resemble \( I_{A}\) which is an image of the letter "A") , \emph{i.e.}:

\begin{multline}
    \mathcal{L}_{\text{A}}\left(\Phi_{1},\Phi_{2}\right) = \\ \frac{1}{H\cdot W} \sum\limits_{n=1}^{H}\sum\limits_{m=1}^{W}{\|I_{mod}\left(n,m\right) -  I_{A}\left(n,m\right)\|_2^2}.
\end{multline}
where \(I_{mod}\left(n,m\right)\) is the resulting diffraction image.

\section{Neural net and training details} \label{SIsec:numerical}

The basic architecture, presented in \cref{fig:SI-NetArc}, of our neural network is similar to the ResNet \cite{he2016deep} network architecture. First, we apply a conv. layer with kernel size of 7x7 pixels to capture features in different sizes and to rapidly increase the receptive field of the next layers. Then we concatenate 7 residual blocks of increasing sizes. Each residual block consists of a skip connection between the input and an additional path through 3 conv. layers (kernel size=3 and padding=1 for all conv. layers) separated by leaky ReLU and batch normalization layers. The final residual block output is passed through two more conv. layers and a log softmax layer that outputs the log of the predicted distributions of the classes in each pixel.

The training set consists of 7000 images and the validation set consists of 2000 images, both chosen randomly from a subset of the BDD100K database. For the focal loss term we have chosen \(\gamma\) to be equal to 2 and \(\alpha_t\) was chosen according to the class frequencies in the training set. We used Adam optimizer with reduce on plateau mechanism (patience=3 and decrease factor=0.1). We have trained the neural network on two Titan RTX GPUs for 48 hours. The network was tested on 1000 images (inference of a single image takes \(\approx\) 10 ms) and the reported accuracy was measured by the number of correctly classified pixels out of the total number of pixels per image.

\section{Additional results} \label{SIsec:numerical}

\begin{figure*}[ht!]
\centering
\includegraphics[scale=1.0]{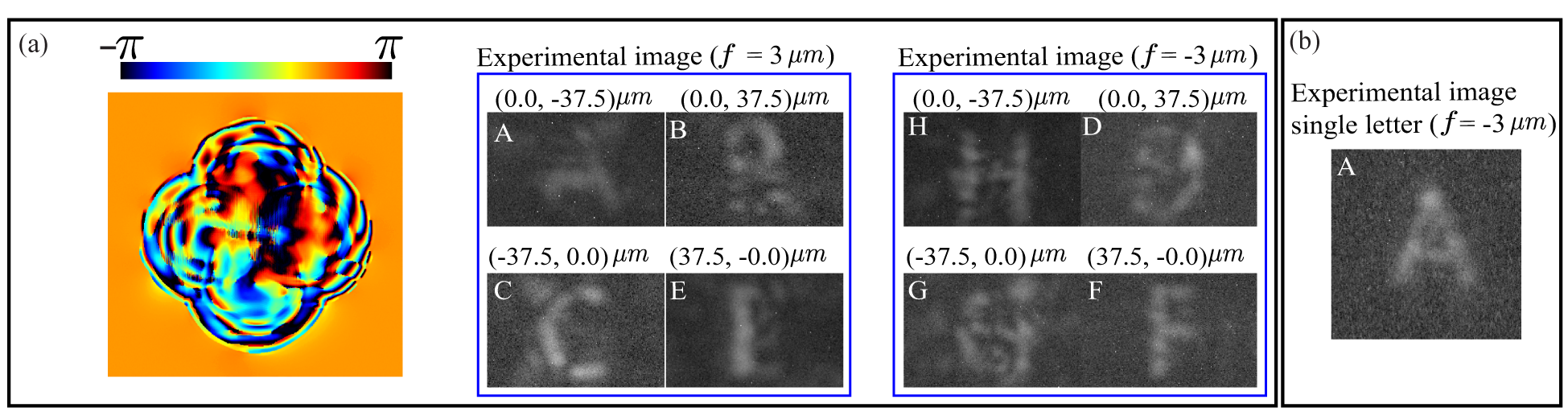}
\caption{Experimental out-of-aperture phase retrieval in 3D. (a) Four lateral position (37.5 \(\mu m\) in each direction from the optical axis) in a two defocus plane (\(\pm\)3 \(\mu m\)). Each position was assigned a letter in the phase retrieval algorithm resulting in the phase mask plotted. The experimental image was acquired by shifting the microscope stage with a sparse bead sample. (b) Experimental result of optimizing a single position (for visual comparison.)}
\label{fig:SI-letters3D}
\end{figure*}

Additionally to performing phase retrieval for a single axial plane in main text fig. 2, we implemented a similar cost function for two axial planes (with eight letters). The result can be seen in \cref{fig:SI-letters3D}. The optimization converged on a less visually satisfying results due to having more constraints to optimize. Nevertheless, this shows the validity of our method to simulate and design spatially-variant PSFs in 3D.

\section{Sampling requirements} 
\label{sec:sampling}

The derivation of the numerical sampling requirements is based on previous research \cite{coy2005choosing,zhang2020analysis}. We denote \(P_{in}\) and \( P_{out} \) as the aperture sizes of the input and output planes sampled with frequencies \(\delta_{in}\) and \( \delta_{out} \), respectively. We also denote \(\mathcal{M}_{j}\) as the ray optics ABCD matrix of the given section. The first two conditions are general for Fourier optics  and they ensure that (from  geometrical considerations in a paraxial regime) the input plane sampling is fine enough to sample the spatial bandwidth and that rays will not wrap around the grid: 

\begin{align}
    \delta_{out} \leqslant -\frac{P_{out}}{P_{in}}\cdot \delta_{in}+\frac{\lambda \cdot B_{j}}{P_{in}},\label{eq:cond_delta1}
    \\
    N \geqslant \frac{P_{in}}{2\delta_{in}} + \frac{P_{out}}{2\delta_{out}}+\frac{\lambda \cdot B_{j}}{2\delta_{in}\cdot\delta_{out}}\label{eq:condN1}
    , 
\end{align}
where \(N\) is the number of grid points (for both dimensions). The next conditions minimize the aliasing of the product of the input field with \( Ch_{in} = \exp\left(i\frac{k}{2\cdot B_{j}}\cdot \left(A_{j}-m\right) \cdot \left|r_{in}\right|^2\right) \). For large apertures, we assume that the highest frequency will be at the edge of the aperture at \( r=P_{in}/2\) due  to the chirp function. However, it is important to note here that in our work we considered defocus but also PSF-engineered wavefronts, so care needs to be taken in each case to avoid aliasing if higher frequencies occur in the  field (the conditions can be simply evaluated numerically for a simulated wavefront). The Nyquist condition for the local frequency at the edge of the input aperture for a input with wavefront curvature \( \frac{k}{2}\cdot S_{U} \): 

 \begin{equation}
 \frac{1}{2\pi}\frac{\partial }{\partial r} \left[\frac{k}{2} \left|S_{U} + \frac{A_{j}-m}{B_{j}} \right| r_{in}^2 \right]_{r_{in}=\frac{P_{in}}{2}}\leqslant \frac{1}{2\delta_{in}}.
 \label{eq:cond-delta-chirp}
 \end{equation}
This equation can be rearranged to two conditions:

\begin{align}
    \delta_{out} \leqslant \left(A_{j}+B_{j}\cdot S_{U}\right)\cdot \delta_{in}+\frac{\lambda\cdot B_{j}}{P_{in}}
    \\
    \delta_{out} \geqslant	\left(A_{j}+B_{j}\cdot S_{U}\right)\cdot \delta_{in}-\frac{\lambda\cdot B_{j}}{P_{in}}.
    \label{eq:sample_cond1&2}
\end{align}
 
\begin{figure}[ht!]
\centering
\includegraphics[scale=1.0]{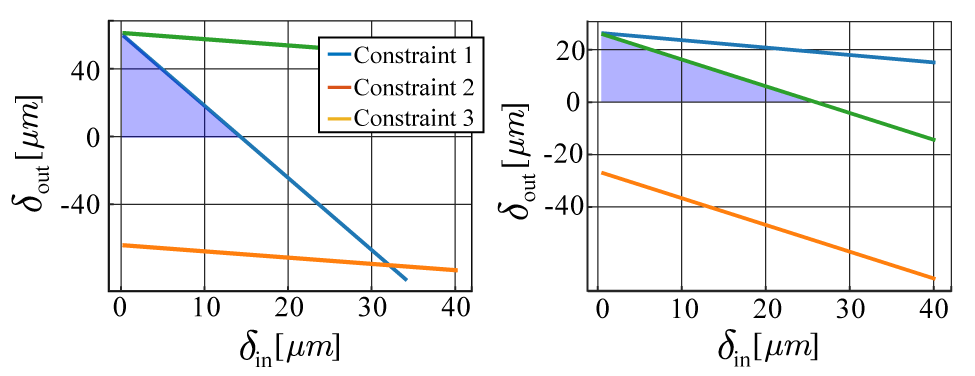}
\caption{Example of the sampling conditions for the experimental system in main text fig.2 (left) Sampling conditions for the first part, (right) sampling conditions for the right part. Blur highlights the viable range. For simplicity, all conditions are calculated for an in-focus PSF, to evaluate the entire system, each focal position needs to be assessed.}
\label{fig:SI-samp}
\end{figure}
The final sampling condition comes from correctly sampling the transfer function \(Ch_{tf} = \exp\left(-i\frac{\pi\lambda B_{j}}{m}\cdot f^2\right) \), where the frequency sampling is \( f=\frac{1}{N\cdot\delta_{in}}\). The Nyquist sampling condition at the edge of the frequency grid (at \(f=\frac{1}{2\delta_{in}}\)) is given by      

\begin{equation}
 \frac{1}{2\pi}\frac{\partial }{\partial f} \left[\frac{\pi\cdot \lambda\cdot B_{j}}{m} f^2 \right]_{f=\frac{1}{2\delta_{in}}}\leqslant \frac{1}{2f} = \frac{N \cdot \delta_{in}}{2}.
 \end{equation}
 This can be simplified to
 
 \begin{equation}
 N \geqslant \frac{\lambda \cdot B_{j}}{\delta_{in}\cdot\delta_{out}}.
 \label{eq:condN2}
 \end{equation}
 The condition for the final output plane chirp can be calculated in a similar  manner to \cref{eq:cond-delta-chirp}. In most of the discussed systems, \(D_{j}\) is a very small number, making this condition extremely lenient. 
 For a simple example, if we look at the experimental system used in main text figs.2 and 3, we plot the possible sampling condition in \cref{fig:SI-samp}. The left plot give the possible sampling range in highlighted blue for the two parts of the system (from point-source to phase mask and from phase mask to camera). The actual choice of the sampling should consider also the two conditions on N, \emph{i.e.} \cref{eq:condN2} and \cref{eq:condN1}, and the desired sampling in the phase mask plane (depending on the desired modulation and used device).  
 In our case, we chose to sample the phase mask with step size of 16 \(\mu m\), which enforced a grid size of N=1024. We note that the experimental system is relatively easy in terms of choosing the sampling conditions compared to the theoretical situation presented in Fig. Main 1, where conditions for three colors are considered. 

\section{Experimental implementation} \label{SIsec:Calibration}

This section provides with details about our sample preparation, optical system (main text fig.2) used and the acquisition of ground truth for the 3D experiment. 

\subsection{Optical system} \label{SIsec:system}

The imaging system (main text fig.2) consists of a Nikon Eclipse-Ti2 inverted fluorescence microscope with a 40X/0.75 NA Nikon objective. An achromatic doublet lens (f=10 cm) images the back focal plane onto an iris. A linear polarizer followed by a LC-SLM (Pluto-VIS130, Holoeye) is placed 10 cm (9.6 cm after calibration- see section \cref{sebsec:reg} ) after the iris. Another achromatic doublet lens (f=20 cm) is placed to form a 4f system. Finally, a sCMOS cameras (PixeLINK PL-D7512MU-T) with pixel size \(3.45\) \(\mu m\) is placed in the image plane. 

\subsection{Sample preparations} \label{SIsec:samples}

For the beads on a coverslip samples ( main text fig. 2 and the calibration described in \cref{subsec:aberr} and \cref{sebsec:reg}) we used a sparse sample consisting of a glass coverslip (170 \(\mu m\), Fisher Scientific) with 200 nm fluorescent beads (Invitrogen, 0.2 \(\mu l\) Crimson 625/645 diluted 1:500) adhered to the surface with 1\% Polyvinyl alcohol (PVA).
For bead immersed in Gel ( main text fig.3) we  used Mini-PROTEAN gel casting stand (BioRad) with 0.75 mm short glass to prepare acrylamide gel with beads by vertical solidification to allow beads to sparsely spread out in 3 dimensions. Gel solution with beads was prepared by gently mixing 1.2 \(ml\) double distilled water, 750 \(\mu l\) 40\% acrylamide/bis-acrylamide solution (Sigma, A7802), 80 \(\mu l\) 50\% glycerol, 3 \(\mu l\) Fluorospheres carboxylate bead solution (Crimson 625/645, Invitrogen), 1.5 \(\mu l\) TEMED, and 10 \(\mu l\) 10\% ammonium per sulphate. The gel solution was immediately pipetted into the gel casting stand and gently topped with about 1 \(ml\) of double distilled water to level the gel. Once solidified the gel was kept in water in the dark at 4oC.

\subsection{Beads in gel ground truth estimation} 

\begin{figure}[h!]
\centering
\includegraphics[scale=1.0]{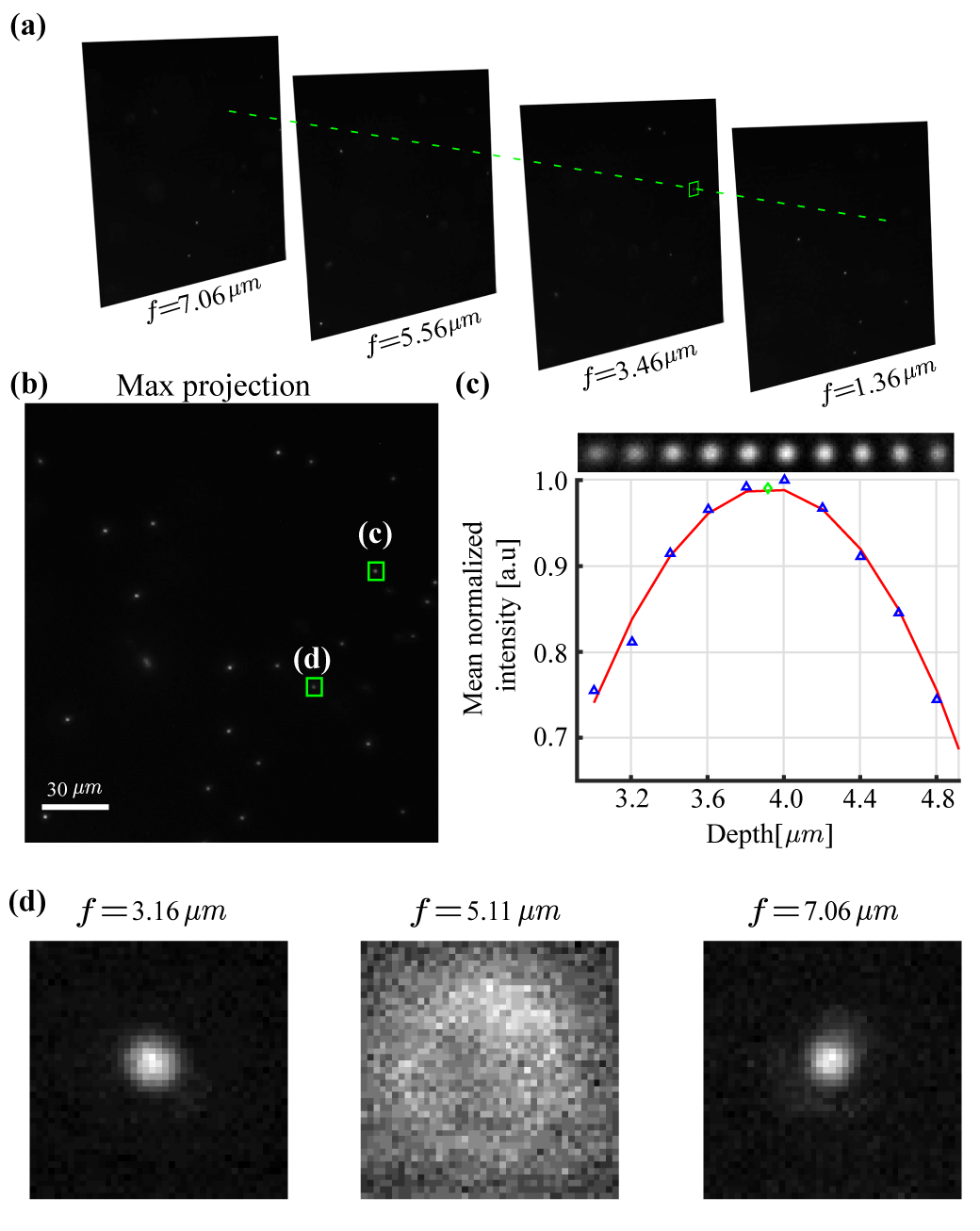}
\caption{Experimental ground truth approximation. (a) The microscope objective is used to create a focal sweep over the axial range of interest (b) Max projection of the captured z-stack (c) Depth fitting example of the mean intensity to determine the axial position of each emitter.(d) Emitter \(\#4\) position where 2 emitters at different axial positions happened to have a very similar lateral position}
\label{figs:exp:gt}
\end{figure}

In the refocusing experiment (main text fig.3), the experimental ground truth 3D positions were approximated \textit{via} a focal scan (\cref{figs:exp:gt}). The scan was performed by moving the objective lens 60 steps of 150 nm each. First, we used ThunderSTORM \cite{ovesny2014thunderstorm} to detect and localize (in 2D) the emitters from the max projection \cref{figs:exp:gt}(a). Next, the focus position per emitter was estimated by fitting a second order polynomial to the mean intensity across focal slices \cref{figs:exp:gt}(c). The mean intensity was estimated by averaging a small ROI of \(7\times7\) pixels around each detected emitter. Finally, the axial position was obtained by correcting the refractive index mismatch between the Gel and air. 

\subsection{Camera to phase mask registration} %
\label{sebsec:reg}
The main spatial calibration that we performed  was to calibrate the LC-SLM axial position and the affine transform between the theoretical and experimental ray tracing of the chief ray from the object plane to the LC-SLM plane. The LC-SLM we used is reflective, thus an angle is introduced which can induce rotations and shearing between the planes. We chose to apply (on the LC-SLM) a phase mask termed the Tetrapod \cite{shechtman2014optimal}. Many optional masks can be used here, but we chose the Tetrapod because a lateral misalignment between the phase mask and input beam can be easily spotted in the z-stack wobble and aberration of the PSF. A sparse bead sample with a single bead in the FOV was used to create a single point source in the image plane. After finding the optical axis on the camera plane, we shifted the mask by 45 LC-SLM pixels on a square grid with 9 points and found the corresponding shifts in the image plane. Finally, an affine transformation was computed between the shifted phase mask grid to the theoretically ray traced positions from the camera plane to the SLM plane. This transformation allowed us to engineer the desired PSF in different FOV positions but also estimate a corrected axial position of the LC-SLM. This was achieved by using a simple ray-tracing for a 4-f system (as plotted in main text fig.2b) with an LC-SLM placed a position d behind the aperture, resulting in the chief ray position \(\left(x_{in},y_{in}\right)\) in the intermediate image plane:

\begin{flalign}
\begin{split}
        x_{SLM} = x_{in}\cdot\left(1-\frac{f-d}{f}\right),\\
        y_{SLM} = y_{in}\cdot\left(1-\frac{f-d}{f}\right).
\end{split}
\end{flalign}

\subsection{Aberration correction} 
\label{subsec:aberr}

In this work, we focused on a microscope with a high NA objective for our experimental validation of the method . Our system does not exhibit noticeable FOV dependant aberrations, and is mainly dominated by shift-invariant spherical and coma aberration. 

We calibrate our system using a method designed for high NA objectives \cite{ferdman2020vipr}. This approach produces a pupil phase correction, as it provides with a good model based on a vectorial diffraction model and inherently (by being pixel-wise) accounts for the issue of wobble.

\bibliographystyle{ieeetr}
\normalsize{\bibliography{references}}

\end{document}